# Interfacial Thermal Transport in Boron Nitride-Polymer Nanocomposite

Ruimin Ma[1], Xiao Wan[3], Teng Zhang[2], Nuo Yang[3]*, Tengfei Luo[2]*

1. School of Energy and Power Engineering, Huazhong University of Science and Technology, Wuhan 430074, China. Current address at Department of Aerospace and Mechanical Engineering, University of Notre Dame, Notre Dame, Indiana 46556, USA

2. Department of Aerospace and Mechanical Engineering, University of Notre Dame, Notre Dame, Indiana 46556, USA.

3. Nano Interface Center for Energy (NICE), School of Energy and Power Engineering, Huazhong University of Science and Technology, Wuhan 430074, China.

*Corresponding author. Email: tluo@nd.edu (T.L); nuo@hust.edu.cn (N.Y)

**ABSTRACT**

Polymer composites with thermally conductive nanoscale filler particles, such as graphene and hexagonal boron nitride (h-BN), are promising for certain heat transfer applications. While graphene-polymer composites have been extensively investigated, studies on h-BN-polymer composites has been relatively rare. In this paper, we use molecular dynamics (MD) simulations to study the interfacial thermal conductance (ITC) involved in the h-BN-polymer composites. We first compare the ITC across h-BN/hexane ($C_6H_{14}$) interfaces to that of graphene/hexane interfaces, where we found that the electrostatic interaction due to the partial charge on h-BN atoms can play an important role in such interfacial thermal transport. Based this finding, we further explore the thermal transport across different h-BN interfaces, including h-BN/hexanamine ($\mathbf{C_6H_{13}NH_2}$), h-BN/hexanol ($\mathbf{C_6H_{13}OH}$), h-BN/hexanoic acid ($\mathbf{C_5H_{11}COOH}$), where the increasingly polar molecules lead to systematic changes in the electrostatic interactions between h-BN and polymers. Heat flux decomposition and atom number density calculations are performed to understand the role of electrostatic interaction in thermal transport across h-BN-polymer interfaces. It was observed that stronger electrostatic interactions across the interfaces can help attract the polymer molecules closer to h-BN, and the reduced interface distance leads to larger heat flux contributed from both van der Waals and electrostatic forces. These results may provide useful information to guide the design of thermally conductive h-BN-polymer nanocomposites.

**INTRODUCTION**

Hexagonal Boron nitride (h-BN) is a layered material similar to graphite with many remarkable properties, such as high in-plane thermal conductivity,[1] wide–band gap,[2] strong oxidation resistance,[3] high tensile strength,[4] efficient light emission[5] and so on.[6,7] Due to these outstanding properties, h-BN has numerous promising applications such as high thermal conductivity insulating materials in electronic devices,[8,9] fillers in high strength and thermally conductive functional composites,[10,11,12] tunneling devices,[13] and field-effect transistors.[14,15] For many of these applications (e.g., h-BN composites for electronics), thermal transport is critical. To achieve high thermal conductivity of h-BN polymer composites, improving interfacial thermal transport between the filler and the polymer matrix can be important, especially when the structure size approaches nanometers and thus the interface density is large.

MD simulations have been widely employed to study the relationship between ITC and interfacial interaction for polymer composites, especially for graphene-polymer composites. In contrast, fundamental studies of the thermal transport between polymer and h-BN, which share a number of similarities (e.g., layered structure, high strength and thermal conductivity) with graphene, are very rare, despite that many experiments have been done to increase the thermal conductivity of h-BN-polymer composite.[19,20,21,22]

In this study, non-equilibrium MD (NEMD) simulations are used to calculate the ITC between h-BN and different polymers. The interfaces are characterized by frequency domain analysis, heat flux decomposition, and atom number density. It is found that when the vibrational modes between the filler and matrix are similar, the ITC will be higher. In addition, the electrostatic forces between h-BN and polymer also pull the polymers closer to the interface, leading to larger van der Waals (vdW) forces, which are very sensitive to interatomic distances. Both the vdW and electrostatic forces contribute to larger heat flux across the interface and thus explain the increasing of thermal conductance. Although fundamental in nature, our studies may provide more ideas on how to increase the thermal conductivity of h-BN nanocomposite. For example, one can use more polar functional groups to modify the polymer matrix to utilize the interfacial electrostatic interactions for better thermal transport.

**SIMULATION METHODS**

We study the interfacial thermal transport in a h-BN based composite. h-BN is used as the filler, and four types of polymer molecules, hexane ($C_6H_{14}$), hexanamine ($\mathbf{C_6H_{13}NH_2}$), hexanol ($\mathbf{C_6H_{13}OH}$), and hexanoic acid ($\mathbf{C_5H_{11}COOH}$) are used as matrices. With the similar carbon backbones and different end groups, the four types of polymers have similar vibrational spectra,[23] but different polarizations. Thus, the differences in ITC of these h-BN/polar polymer systems can be attributed to polarization.

The Tersoff potential[24][25][26] is used to describe the interaction between boron nitride atoms, and universal force field (UFF)[27] is used for the non-bonding interaction between h-BN and organic molecules. The polymers are simulated using the polymer consistent force field (PCFF)[28]. All simulations are carried out using the large-scale atomic/molecular massively parallel simulator (LAMMPS).[29] A timestep size of 0.25 fs is used for all simulations.

A typical structure is shown in **Figure 1 (a)**. A single layer of h-BN is centered in the simulation domain with periodic boundary conditions (PBCs) in all three directions. The PBCs in the lateral direction making the h-BN effective infinite in size without edges. This is reasonable since majority of the interfaces are between the basal plane of h-BN and the polymer matrices. The paraffin wax system contains 100 polymer molecules, and the three polar polymer systems each contain 200 organic molecules. The whole system is heated up to 600K under NPT (constant number of atoms, pressure and temperature) ensemble to achieve the disordered amorphous phase, and then cooled down to 300K with an annealing speed of 12 K/ps.

NEMD is used to calculate the interfacial thermal conductance. After NPT (constant number of atoms, pressure and temperature) relaxation at 300 K and 1 atm for 1.5-2 ns, the system is then run in an NVE (constant number of atoms, volume and energy) ensemble with temperatures of the heat source and sink regions at the ends of the system set to 320K and 280K, respectively, using Langevin thermostats (Figure 1a). A layer of atoms at each end of the simulation domain is fixed to prevent the heat transfer across the periodic boundaries, forcing all heat flux to across the h-BN-polymer interface. The fixed atoms also prevent the translational drift of the whole system, and thus help 'lock' the position of the interface and extract the temperature profile. The relatively large temperature difference (40 K) is used to establish a measurable temperature gap across the interface. Except the fixed atoms, the rest of system, including the thermostated regions, is simulated in the NVE ensemble for 10 ns to reach the steady state and for ITC calculations. A typical temperature profile at the steady state is shown in **Figure 1 (b)**. The heat flux (*q*) is calculated by averaging the energy input and output rates from the heat source and sink. The temperature jump *(ΔT)* across the interface is defined by the temperature difference between polymer atoms at the two sides of the h-BN layer. Interfacial thermal conductance (*G*) is then calculated as $G = q/\Delta T$. The calculated *G* is thus from two h-BN-polymer interfaces.

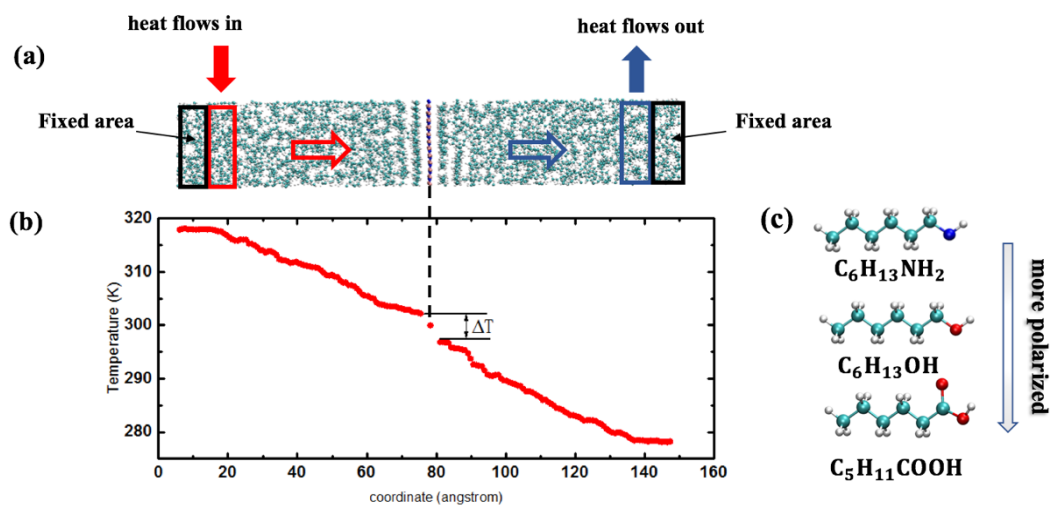

**Figure 1.** (a) Simulation setup for ITC calculations in the NEMD method: heat flows across the interface from source to sink with the black areas being the fixed regions. (b) Steady state temperature profile of the simulation cell and the temperature difference across the interface. (c) The structures of three different kinds of polar materials.

**RESULTS AND DISCUSSION**

**ITC comparison between graphene/polyethylene and h-BN/polyethylene interfaces.** ITC of the h-BN/polyethylene interface is found to be $131 \pm 36$ MW/m$^2$-K, which is larger than that of graphene/polyethylene interface (61-71 MW/m$^2$-K). From a microscopic point of view, thermal transport across an interface is accomplished through the interaction between the two groups of atoms across the interface. Stronger interaction has been proven to enable larger ITC. The non-bonded interactions between polyethylene and h-BN or graphene are vdW forces, since we only consider the vdW forces at the start. The vdW interaction in our model is described by a 12/6 Lennard-Jones (L-J) potential, and h-BN is found to have larger energy constants than graphene (**Table 1**),[27][30] indicating the stronger interactions and explaining the larger interfacial thermal conductance between h-BN and paraffin.

Table 1. L-J potential energy parameters for different atom species.

| Atom type | Energy constant ($\varepsilon$) [eV] |
|---|---|
| C (graphene) | 0.002390 |
| B (boron nitride) | 0.006000 |
| N (boron nitride) | 0.003700 |

Another important factor for thermal transport between two materials is the matching of the vibration power spectra (VPS) in the frequency domain, which can be calculated by applying Fourier transform on the velocity autocorrelation functions (VAF) of atoms.

$$D(\omega) = \int_0^\tau \Gamma(t)\cos(\omega t)dt$$

$$\Gamma(t) = <v(t)v(0)>$$

where $\omega$ is frequency, $D(\omega)$ is the VPS at frequency $\omega$, and $\Gamma(t)$ is the VAF of atoms.

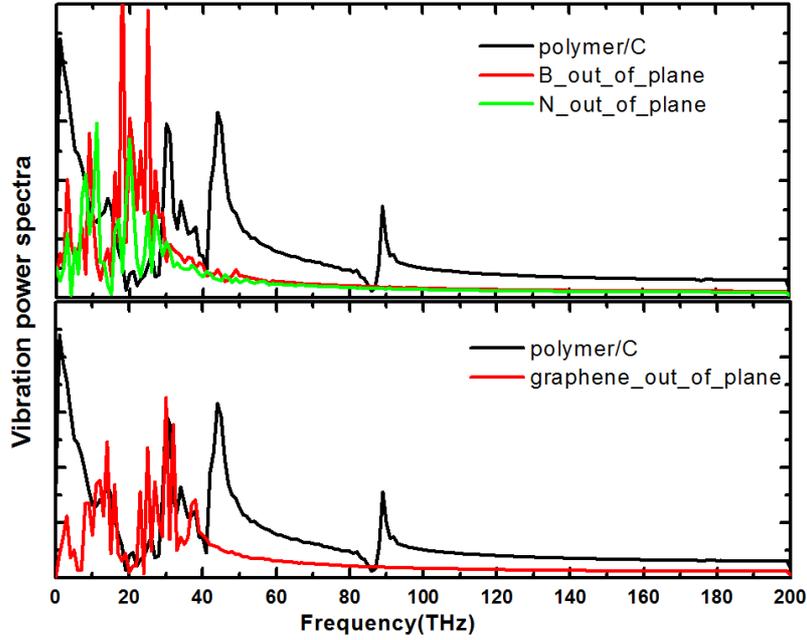

**Figure 2.** The vibration power spectra of graphene/paraffin wax system and h-BN/paraffin wax system.

In a previous work,[30] the out-of-plane motions of graphene atoms are found to couple to polymer atom motions better. Therefore, here we only show the out-of-plane components of both graphene and h-BN vibrational modes (**Figure 2**; in-plane motions of h-BN in Supporting Information). In another study, Hu et al.[31] also found that the coupling between graphene out-of-plane motion and polymer molecule motions at low frequencies serves as the most important channels for thermal transport across the interface. We can see from Figure 2 that in the low-frequency region (0-20 THz), h-BN spectrum is apparently coupling much better with polymer spectrum, providing another reason for the higher ITC of h-BN/polyethylene interface.

**The impact of electrostatic interaction on the ITC of h-BN/polyethylene interfaces.** To study the impact of Coulombic interaction, we first do an artificial simulation by removing the partial charges assigned to the atoms of polyethylene. **Table 2** shows the results of ITC with and without Coulombic contribution. From the results, we can see that there is no obvious difference between them. Since the polarity of polyethylene is very weak, the Coulombic contribution can be neglected.

Table 2. ITC with and without Coulombic contribution.

| h-BN/polyethylene system | ITC [MW/m^2-K] |
|---|---|
| With coulomb contribution | **142 ± 27** |
| Without coulomb contribution | **131 ± 36** |

**The impact of electrostatic interaction on the ITC of h-BN/polar polymer interfaces.** We calculate the ITC of h-BN/hexanamine, h-BN/hexanol and h-BN/hexanoic acid under the cases with and without Coulombic contribution. As shown in **Figure 4**, opening coulomb channel in the system will improve the ITC, as the electrostatic force will strengthen the interfacial interaction. When the materials become more polarized, the electrostatic force become stronger and thus lead to even larger ITC.

To isolate the contributions from different types of interactions, we decompose the total heat flux across the interface into the electrostatic and vdW parts. The power exchange between the h-BN group and polymer group can be calculated as:[32,33]

$$p = \frac{1}{2}(\sum_{i \in BN} F_{ij} v_j - \sum_{i \in polymer} F_{ji} v_i)$$

where $F_{ij}$ is the force on atom $j$ from atom I and $v_j$ is the velocity of atom $j$. The total force at the interface can be decomposed into the vdW force and the electrostatic force. Thus, the above power exchange function can be written in terms of these two types of forces:

$$p = (p^{vdW} + p^q)$$

$$= \frac{1}{2}\left(\sum_{i \in BN} F_{ij}^{vdW} v_j - \sum_{i \in polymer} F_{ji}^{vdW} v_i\right) + \frac{1}{2}\left(\sum_{i \in BN} F_{ij}^{q} v_j - \sum_{i \in polymer} F_{ji}^{q} v_i\right)$$

Figure 5 shows the time integration of power exchange across three representative interfaces, in other words, cumulative energies across interfaces. The red lines ($p^q$) show the contribution from electrostatic force in total heat flux. From the least polarized system (h-BN/$C_6H_{15}N$) to the most polarized system (h-BN/$C_5H_{11}COOH$), $p^q$ significantly increases, leading to a higher ITC. Comparing the h-BN/$C_6H_{13}OH$ system and h-BN/ $C_5H_{11}COOH$ system, we also found that the contribution from electrostatic force to the heat flux is almost the same; but the contribution from vdW force to heat flux is much larger in h-BN/ $C_5H_{11}COOH$ system. So, the vdW forces increases rapidly as the electrostatic force increases in the system.

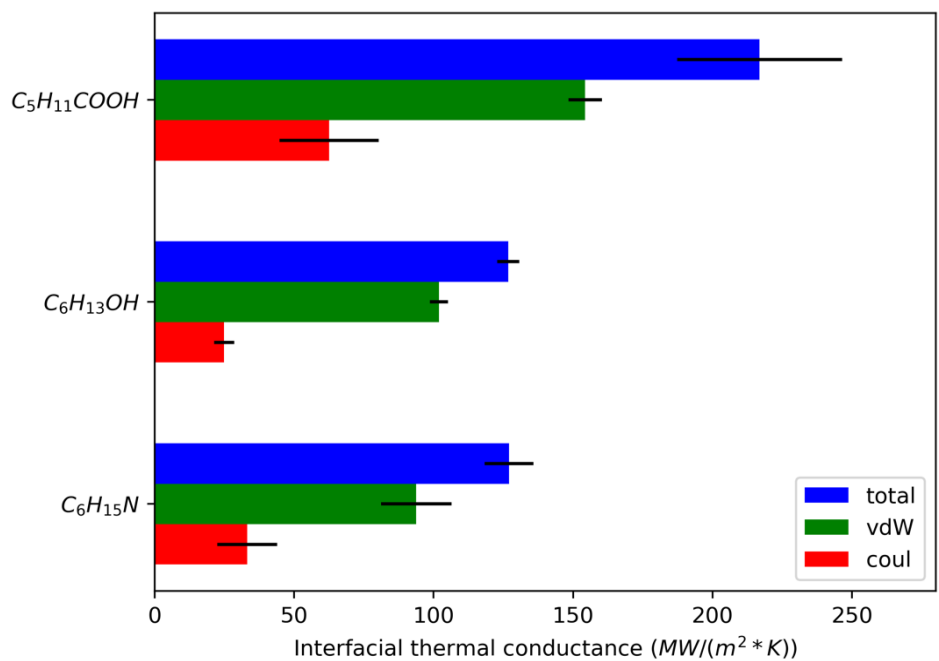

**Figure 4.** The interfacial conductance of h-BN/ hexanamine, h-BN/ hexanol and h-BN/hexanoic acid counted from the contribution of different forces. The red bars illustrate the ITC counted from the contribution of electrostatic force; the green bars illustrate the ITC counted from the vdW force; and the blue bars illustrate the ITC counted from the contribution from both the vdW and electrostatic forces.

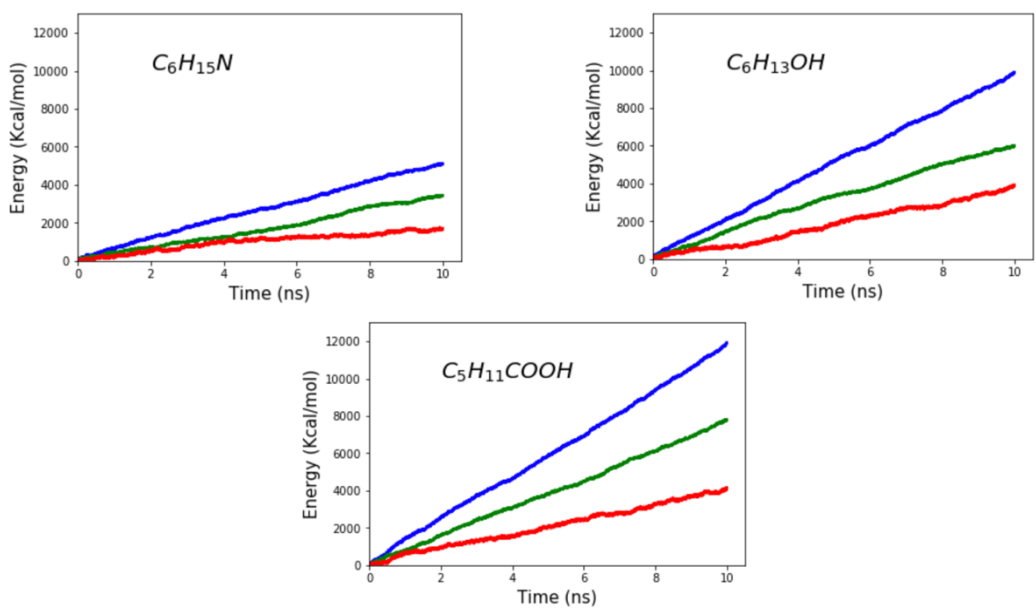

**Figure 5.** Heat flux decomposition on three representative interfaces: h-BN/$C_6H_{15}N$, h-BN/$C_6H_{13}OH$, h-BN/$C_5H_{11}COOH$. The red lines in the three graphs above indicate the contribution from electrostatic force to the heat flux; the green lines indicate the contribution from vdW force to the heat flux; and the blue lines indicate the overall heat flux.

**Atom number density analysis for explaining the effect of electrostatic interaction**. We calculate the box volumes in three representative systems (See **Table 3**), we can see that the box volume in the open-coulomb-channel case is smaller than that in the close-coulomb-channel case. It is because that the electrostatic force will pull atoms closer to each other, which makes the system more compact, as we open the coulomb channel in the system. Such an effect can be visualized via atom number density (See **Figure 6**). From the figure, we can see that the density peak shift closer to the interface when we open the coulomb channel in the system, which pushes more atoms to join the energy transferring process near the interface. The ITC then becomes larger.

Table 3. Box sizes along the heat flux direction in three representative systems.

| System | Case | Box Volume (Angstrom^3) |
|---|---|---|
| h-BN/$C_6H_{15}N$ | Open coulomb channel | 56677 |
| | Close coulomb channel | 63286 |
| h-BN/$C_6H_{13}OH$ | Open coulomb channel | 47426 |
| | Close coulomb channel | 52523 |
| h-BN/$C_5H_{11}COOH$ | Open coulomb channel | 44494 |
| | Close coulomb channel | 47436 |

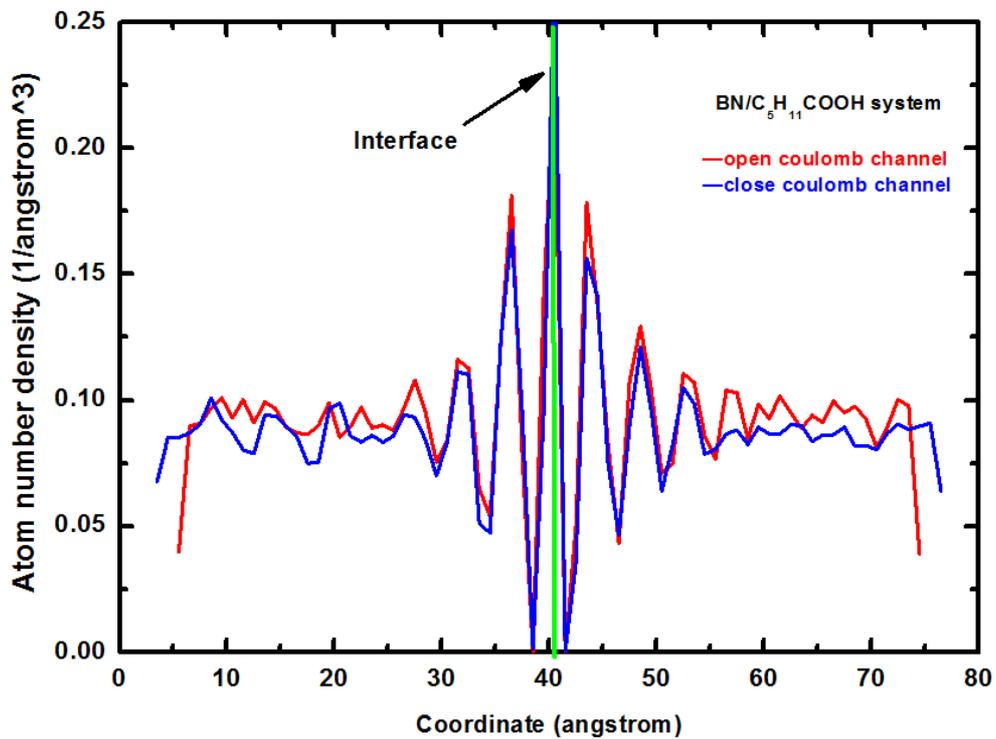

Figure 6. The atom number density of the h-BN/$C_5H_{11}COOH$ system. The green line in the middle is the position of BN. The red line shows the atom number density of the system with coulomb channel opened; the blue line shows the atom number density of the system with coulomb channel closed.

## CONCLUSION

In conclusion, we use the steady-state NEMD method to calculate the ITC in h-BN based nanocomposite. Comparison with the ITC of graphene/paraffin wax system, h-BN is more effective than graphene in terms of interfacial thermal transport due to the larger energy constants and similar vibrational frequency. L-J parametric study is used to emphasize the importance of interfacial interaction. From the investigation of the thermal transport in h-BN/polar materials system, electrostatic interactions are found to play a critical role in ITC. More polarized the materials have larger the ITC. Through the heat flux decomposition and atom number density analysis, electrostatic interactions compacts the system and induces larger vdW interactions. All in all, our results provide guidance in improving the thermal conductivity of h-BN based nanocomposite.